\documentclass[10pt]{article}
\usepackage{graphicx}
\usepackage{amsmath}
\usepackage{amssymb}
\usepackage{caption2}
\usepackage{slashed}
\begin{document}
\bibliographystyle{prsty}
\begin{center}
{\large {\bf \sc{Strong decays of the fully-charm tetraquark states with explicit P-waves via the QCD sum rules}}} \\[2mm]
Xiao-Song Yang$^{*\dag}$, Zhi-Gang Wang$^*$\footnote{E-mail: zgwang@aliyun.com.  }    \\
Department of Physics, North China Electric Power University, Baoding 071003, P. R. China$^*$
School of Nuclear Science and Engineering, North China Electric Power University, Beijing 102206, P. R. China$^\dag$
\end{center}

\begin{abstract}
We introduce a relative P-wave to construct the vector doubly-charm diquark $(\widetilde{V})$, therefore,  the scalar and tensor four-quark currents to investigate the decay widths of the fully-charm tetraquark states with the $J^{PC}=0^{++}$, $1^{+-}$ and $2^{++}$ via the QCD sum rules.
We observe that the total width of the ground state $\widetilde{V}\overline{\widetilde{V}}$-type scalar tetraquark state is compatible with that of the $X(6552)$ within the uncertainties, and the branching ratios are quite different from that of the first radial excitation of the $A\bar{A}$-type scalar  tetraquark state.
Other predictions can be verified in the future experiments to shed light on the nature of the fully-charm tetraquark states.
\end{abstract}



\section{Introduction}
 Recently, the fully heavy hadrons  become a research hot-spot in high energy physics, while the first fully-heavy tetraquark candidates were reported by the LHCb collaboration in 2020 \cite{LHCb-cccc-2006}.
The LHCb collaboration  observed a narrow structure $X(6900)$ and a broad structure above the di-$J/\psi$ threshold ranging from 6.2 to 6.8 GeV in the $J/\psi J/\psi$  invariant mass spectrum using the proton-proton collision data at $\sqrt{s}=7$, $8$ and $13\,\rm{TeV}$, which corresponds to an integrated luminosity of $9\, \rm{fb}^{-1}$ \cite{LHCb-cccc-2006}.

Subsequently, the ATLAS collaboration confirmed the $X(6900)$ and observed several resonances ($R$) in the $J/\psi J/\psi$ and $J/\psi\psi^\prime$ invariant mass spectra based on the proton-proton collision data at $\sqrt{s}=13\,\rm{TeV}$ corresponding to an integrated luminosity of $140\, \rm{fb}^{-1}$ in 2023 \cite{ATLAS-cccc-2023}.
The fitted Breit-Wigner masses and widths are given as follows,
\begin{flalign}
 & R_0 : M = 6.41\pm 0.08_{-0.03}^{+0.08} \mbox{ GeV}\, , \, \Gamma = 0.59\pm 0.35_{-0.20}^{+0.12} \mbox{ GeV} \, , \nonumber\\
 & R_1 : M = 6.63\pm 0.05_{-0.01}^{+0.08} \mbox{ GeV}\, , \, \Gamma = 0.35\pm 0.11_{-0.04}^{+0.11} \mbox{ GeV} \, , \nonumber\\
 & R_2 : M = 6.86\pm 0.03_{-0.02}^{+0.01} \mbox{ GeV}\, , \, \Gamma = 0.11\pm 0.05_{-0.01}^{+0.02} \mbox{ GeV} \, ,
\end{flalign}
or
\begin{flalign}
 & R_0 : M = 6.65\pm 0.02_{-0.02}^{+0.03} \mbox{ GeV}\, , \, \Gamma = 0.44\pm 0.05_{-0.05}^{+0.06} \mbox{ GeV} \, , \nonumber\\
 & R_2 : M = 6.91\pm 0.01\pm 0.01 \mbox{ GeV}\, , \, \Gamma = 0.15\pm 0.03\pm 0.01 \mbox{ GeV} \, ,
\end{flalign}
in the di-$J/\psi$ mass spectrum, and
\begin{flalign}
 & R_3 : M = 7.22\pm 0.03_{-0.04}^{+0.01} \mbox{ GeV}\, , \, \Gamma = 0.09\pm 0.06_{-0.05}^{+0.06} \mbox{ GeV} \, ,
\end{flalign}
or
\begin{flalign}
 & R_3 : M = 6.96\pm 0.05\pm 0.03 \mbox{ GeV}\, , \, \Gamma = 0.51\pm 0.17_{-0.10}^{+0.11} \mbox{ GeV} \, ,
\end{flalign}
in the $J/\psi \psi^\prime$ mass spectrum.

In the same year, the CMS collaboration studied the $J/\psi J/\psi$ invariant mass spectrum produced in the proton-proton collisions at center-of-mass energy of $\sqrt{s} = 13\,\rm{TeV}$, which corresponds to an integrated luminosity of $135\,\rm{ fb}^{-1}$ \cite{CMS-cccc-2023}. In this study, they observed three resonant  structures with the fitted Breit-Wigner masses and widths of
\begin{flalign}
 & R_1 : M = 6552\pm10\pm12 \mbox{ MeV}\, , \, \Gamma = 124^{+32}_{-26}\pm33 \mbox{ MeV} \, , \nonumber\\
 & R_2 : M = 6927\pm9\pm4 \mbox{ MeV}\, , \, \Gamma = 122^{+24}_{-21}\pm18 \mbox{ MeV} \, , \nonumber\\
 & R_3 : M = 7287^{+20}_{-18}\pm5 \mbox{ MeV}\, , \, \Gamma = 95^{+59}_{-40}\pm19 \mbox{ MeV} \, ,
\end{flalign}
and local significance of $6.5$, $9.4$ and $4.1$ standard deviations, respectively.

The quantum numbers $J^{PC}$ of those newly observed resonances have not been determined until now and their inner structures are still under hot debate.
On the theoretical side, the fully-charm tetraquark states were investigated by several  phenomenological approaches, such as the potential quark  model \cite{Rosner-2017,Wu-2018,FKGuo-2018-Anwar,Polosa-2018,Navarra-2019,Bai-2016,Zhong-2019,
Roberts-2020,PingJL-2020,Zhong-2021,Mutuk-2021,DWC-2023-PRD,YGL-2023-EPJC,Jia-2023,Ping-2024}, the QCD sum rules \cite{WZG-cccc-EPJC,Chen-2017,WZG-cccc-APPB,WZG-cccc-CPC,WZG-cccc-IJMPA,ZhangJR-PRD,
QiaoCF-2021,WZG-cccc-NPB,X6600-Azizi,X6900-Azizi,WZG-YXS-cccc-AAPPS,Tang-2024,Chen-2024}, the lattice QCD \cite{Hughes-2018}, the dynamical rescattering mechanism \cite{WangJZ-Produ-mass}, the Bethe-Salpeter (BS) equation \cite{Zhu-2021-NPB}, and the coupled-channel final state interactions \cite{GuoFK-2021-PRL,XKDong-SB-2021,Gong-Zhong-2022}. Nevertheless, none of them can explain all the resonances consistently and we still need more experimental data to figure out the nature of the fully-charm tetraquark states unambiguously.

In our previous studies, we studied the mass spectrum of the ground state and first radial excited tetraquark states (which are constructed by the axialvector diquarks  $\varepsilon^{ijk}Q_j^T C\gamma_\mu Q_k$ $(A)$) with the spin-parity-charge-conjugation  $J^{PC}=0^{++}$, $1^{+-}$, $1^{--}$ and $2^{++}$ \cite{WZG-cccc-EPJC,WZG-cccc-APPB,WZG-cccc-CPC}. In Ref.\cite{WZG-cccc-NPB}, we considered the updated experimental data and re-studied the mass spectrum of the ground, first, second and third radial excited $A\bar{A}$-type fully-charm tetraquark states with the spin-parity-charge-conjugation $J^{PC}=0^{++}$, $1^{+-}$ and $2^{++}$. Subsequently, we extended this work to explore the strong decays of the ground states and first radial excited tetraquark states via the QCD sum rules \cite{WZG-YXS-cccc-AAPPS}. Combined with the masses and decay widths, we can come to a conclusion that the $X(6552)$ can be assigned as the first radial excitation of the $A\bar{A}$-type scalar  tetraquark state.

In Ref.\cite{WZG-cccc-IJMPA}, we introduced a relative P-wave to construct the doubly-charm vector diquarks $\varepsilon^{ijk}Q^T_j C\gamma_5 \overset\leftrightarrow{\partial}_\mu Q_k$ $(\widetilde{V})$ and constructed the $\widetilde{V}\overline{\widetilde{V}}$-type tetraquark currents to study the mass spectrum of the ground state fully-charm tetraquark states with the spin-parity-charge-conjugation  $J^{PC}=0^{++}$, $1^{+-}$ and $2^{++}$ via the QCD sum rules. The numerical results indicate that the ground state $\widetilde{V}\overline{\widetilde{V}}$-type tetraquark states and the first radial excited $A\bar{A}$-type tetraquark states have almost degenerated masses.

As the assignments by the masses alone are imprecise, in the present work, we explore the decay widths of the scalar, axialvector and tensor $\widetilde{V}\overline{\widetilde{V}}$-type tetraquark states in the framework of the QCD sum rules, which works well in several works on the hadronic coupling constants and decay widths \cite{WZG-ZJX-Zc-Decay,WZG-Y4660-Decay,WZG-X4140-decay,WZG-Zcs3985-decay,WZG-Zcs4123-decay,
WZG-Z4500-LC-decay,WZG-3872-decay,WZG-Y4500-decay}, and make more credible assignments based on the masses and widths together to diagnose the nature of the fully-charm tetraquark states.

The article is organized as follows: in Section 2, we obtain the hadronic coupling constants of the $\widetilde{V}\overline{\widetilde{V}}$-type tetraquark states for seven decay channels; in Section 3, we make reasonable discussions for the numerical results; and finally, the conclusion of this study is presented in Section 4.

\section{ QCD sum rules for the hadronic coupling constants }

The fully-charm interpolating currents with two P-waves are constructed as,
\begin{eqnarray}
J^0(x) &=& \varepsilon^{ijk} \varepsilon^{imn} c^T_j(x) C \gamma_5 \overset\leftrightarrow{\partial}_{\alpha} c_k(x) \bar{c}_m(x) \overset\leftrightarrow{\partial}_{\beta} \gamma_5 C \bar{c}^T_n(x) g^{\alpha\beta}\, , \nonumber\\
J_{\alpha\beta}^1(x) &=& \varepsilon^{ijk} \varepsilon^{imn} \Big\{c^T_j(x) C \gamma_5 \overset\leftrightarrow{\partial}_{\alpha} c_k(x) \bar{c}_m(x) \overset\leftrightarrow{\partial}_{\beta} \gamma_5 C \bar{c}^T_n(x) \nonumber\\
&& -c^T_j(x) C \gamma_5 \overset\leftrightarrow{\partial}_{\beta} c_k(x) \bar{c}_m(x) \overset\leftrightarrow{\partial}_{\alpha} \gamma_5 C \bar{c}^T_n(x) \Big\}\, ,\nonumber\\
J_{\alpha\beta}^2(x) &=& \varepsilon^{ijk} \varepsilon^{imn} \Big\{c^T_j(x) C \gamma_5 \overset\leftrightarrow{\partial}_{\alpha} c_k(x) \bar{c}_m(x)\overset\leftrightarrow{\partial}_{\beta} \gamma_5 C \bar{c}^T_n(x) \nonumber\\
&&+c^T_j(x) C \gamma_5 \overset\leftrightarrow{\partial}_{\beta} c_k(x) \bar{c}_m(x) \overset\leftrightarrow{\partial}_{\alpha} \gamma_5 C \bar{c}^T_n(x) \Big\}\, ,
\end{eqnarray}
where the $i$, $j$, $k$, $m$ and $n$ are color indexes \cite{WZG-cccc-IJMPA}, and the currents,
\begin{eqnarray}
J^{\eta_c}(x) &=& \bar{c}(x)i\gamma_5 c(x)\, , \nonumber\\
J^{J/\psi}_{\alpha}(x) &=& \bar{c}(x)\gamma_\alpha c(x)\, , \nonumber\\
J^{\chi_c}_{\alpha}(x) &=& \bar{c}(x)\gamma_\alpha \gamma_5 c(x)\, , \nonumber\\
J^{h_c}_{\alpha\beta}(x) &=& \bar{c}(x)\sigma_{\alpha\beta} c(x)\, ,
\end{eqnarray}
interpolate the conventional mesons $\eta_c$, $J/\psi$, $\chi_{c1}$ and $h_c$, respectively.

Based on above currents, we adopt the three-point correlation functions,
\begin{eqnarray}\label{CF-0}
\Pi^1(p,q) &=& i^2 \int d^4 x d^4 y e^{ip\cdot x} e^{iq\cdot y} \langle0| T\left\{J^{\eta_c}(x) J^{\eta_c}(y) J^{0\dagger}(0)\right\}|0\rangle \, , \nonumber\\
\Pi^2_{\alpha\beta}(p,q) &=& i^2 \int d^4 x d^4 y e^{ip\cdot x} e^{iq\cdot y} \langle0| T\left\{J^{J/\psi}_{\alpha}(x) J^{J/\psi}_{\beta}(y) J^{0\dagger}(0)\right\}|0\rangle \, , \nonumber\\
\Pi^3_{\alpha}(p,q) &=& i^2 \int d^4 x d^4 y e^{ip\cdot x} e^{iq\cdot y} \langle0| T\left\{J^{\chi_c}_{\alpha}(x) J^{\eta_c}(y) J^{0\dagger}(0)\right\}|0\rangle \, ,
\end{eqnarray}
\begin{eqnarray}\label{CF-1}
\Pi^4_{\mu\alpha\beta}(p,q) &=& i^2 \int d^4 x d^4 y e^{ip\cdot x} e^{iq\cdot y} \langle0| T\left\{J^{J/\psi}_{\mu}(x) J^{\eta_c}(y) J_{\alpha\beta}^{1\dagger}(0)\right\}|0\rangle \, , \nonumber\\
\Pi^5_{\mu\nu\alpha\beta}(p,q) &=& i^2 \int d^4 x d^4 y e^{ip\cdot x} e^{iq\cdot y} \langle0|
T\left\{J^{h_c}_{\mu\nu}(x) J^{\eta_c}(y) J_{\alpha\beta}^{1\dagger}(0)\right\}|0\rangle \, ,
\end{eqnarray}
\begin{eqnarray}\label{CF-2}
\Pi^6_{\alpha\beta}(p,q) &=& i^2 \int d^4 x d^4 y e^{ip\cdot x} e^{iq\cdot y} \langle0| T\left\{J^{\eta_c}(x) J^{\eta_c}(y) J_{\alpha\beta}^{2\dagger}(0) \right\}|0\rangle \, , \nonumber\\
\Pi^7_{\mu\nu\alpha\beta}(p,q) &=& i^2 \int d^4 x d^4 y e^{ip\cdot x} e^{iq\cdot y} \langle0| T\left\{J^{J/\psi}_{\mu}(x) J^{J/\psi}_{\nu}(y) J_{\alpha\beta}^{2\dagger}(0)\right\}|0\rangle \, ,
\end{eqnarray}
to study the hadronic coupling constants, and therefore the widths of the decay channels,
\begin{eqnarray}
X_0 &\to& \eta_c + \eta_c\, , \nonumber\\
X_0 &\to& J/\psi + J/\psi\, , \nonumber\\
X_0 &\to& \chi_c + \eta_c\, , \nonumber\\
X_1 &\to& J/\psi + \eta_c\, , \nonumber\\
X_1 &\to& h_c + \eta_c\, , \nonumber\\
X_2 &\to& \eta_c + \eta_c\, , \nonumber\\
X_2 &\to& J/\psi + J/\psi\, ,
\end{eqnarray}
where the subscripts denote the spins $0$, $1$ and $2$, respectively.
In the present work, we choose the supposed dominant decays, which take place through the Okubo-Zweig-Iizuka super-allowed fall-apart mechanism, and ignore the tiny contributions of the other non-dominant (Okubo-Zweig-Iizuka suppressed) decays approximately, such as the decays $X_0 \to D\bar{D}$ or $D^*\bar{D}^*$, which can occur by annihilating a $c\bar{c}$ pair and creating a $q\bar{q}$ pair.

In the next step, we insert complete sets of intermediate hadronic states with the same quantum numbers as the currents into the hadron side of those  correlation functions  \cite{SVZ79,Reinders85}, and write down the explicit expressions of the ground state contributions (isolated in the charmonium channels),
\begin{eqnarray}\label{Hadron-CT-1}
\Pi^1(p,q)&=& \frac{\lambda_{X_0} f_{\eta_c}^2 m_{\eta_c}^4 G_{X_0\eta_c \eta_c} }{4m_c^2(m_{X_0}^2-p^{\prime2})(m_{\eta_c}^2-p^2)(m_{\eta_c}^2-q^2)}   + \cdots\, , \nonumber\\
&=&\Pi_{1}(p^{\prime2},p^2,q^2)  + \cdots\, ,
\end{eqnarray}

\begin{eqnarray}\label{Hadron-CT-2}
\Pi^2_{\alpha\beta}(p,q)&=& \frac{\lambda_{X_0} f_{J/\psi}^2 m_{J/\psi}^2 G_{X_0J/\psi J/\psi} }{(m_{X_0}^2-p^{\prime2})(m_{J/\psi}^2-p^2)(m_{J/\psi}^2-q^2)}\,  g_{\alpha\beta}  + \cdots\, , \nonumber\\
&=&\Pi_{2}(p^{\prime2},p^2,q^2)\,   g_{\alpha\beta}  + \cdots\, ,
\end{eqnarray}

\begin{eqnarray}\label{Hadron-CT-3}
\Pi^3_{\alpha}(p,q)&=& -\frac{\lambda_{X_0} f_{\chi_c} m_{\chi_c} f_{\eta_c}m_{\eta_c}^2 G_{X_0\chi_c \eta_c}   }{2m_c(m_{X_0}^2-p^{\prime2})(m_{\chi_c}^2-p^2)(m_{\eta_c}^2-q^2)}\,iq_\alpha  + \cdots\, , \nonumber\\
&=&\Pi_{3}(p^{\prime2},p^2,q^2)\, (-iq_\alpha)    + \cdots\, ,
\end{eqnarray}

\begin{eqnarray}\label{Hadron-CT-4-project}
P_{A}^{\alpha\beta\alpha^\prime\beta^\prime}(p^\prime)\,\epsilon_{\alpha^\prime\beta^\prime}{}^{\mu\tau}\,p_\tau\,\Pi^4_{\mu\alpha\beta}(p,q)&=&
\widetilde{\Pi}_4(p^{\prime2},p^2,q^2)\left(p^2+p\cdot q\right)\, ,
\end{eqnarray}

\begin{eqnarray}\label{Hadron-CT-4}
\Pi_4(p^{\prime2},p^2,q^2)&=&\widetilde{\Pi}_4(p^{\prime2},p^2,q^2)\,p^2\, ,\nonumber\\
&=&\frac{\lambda_{X_1} f_{J/\psi}m^3_{J/\psi}f_{\eta_c} m_{\eta_c}^2 G_{X_1J/\psi \eta_c} }{2m_cm_{X_1}(m_{X_1}^2-p^{\prime2})(m_{J/\psi}^2-p^2)(m_{\eta_c}^2-q^2)}  + \cdots\, ,
\end{eqnarray}

\begin{eqnarray}\label{Hadron-CT-5-project}
P_{A}^{\mu\nu\mu^\prime\nu^\prime}(p)P_{A}^{\alpha\beta\alpha^\prime\beta^\prime}(p^\prime)\,\epsilon_{\mu^\prime\nu^\prime\alpha^\prime\beta^\prime}\,
\Pi^5_{\mu\nu\alpha\beta}(p,q)&=&i\,
\widetilde{\Pi}_5(p^{\prime2},p^2,q^2)\left(-p^2q^2+(p\cdot q)^2\right)\,p\cdot q\, ,
\end{eqnarray}

\begin{eqnarray}\label{Hadron-CT-5}
\Pi_5(p^{\prime2},p^2,q^2)&=&\widetilde{\Pi}_5(p^{\prime2},p^2,q^2)\,p^2q^2 \,p\cdot q \, ,\nonumber\\
&=&\frac{\lambda_{X_1} f_{h_c}m^2_{h_c}f_{\eta_c} m_{\eta_c}^4 G_{X_1h_c \eta_c} }{9m_cm_{X_1}(m_{X_1}^2-p^{\prime2})(m_{h_c}^2-p^2)(m_{\eta_c}^2-q^2)}\,p\cdot q + \cdots\, ,
\end{eqnarray}

\begin{eqnarray}\label{Hadron-CT-6}
\Pi^6_{\alpha\beta}(p,q)&=&- \frac{\lambda_{X_2} f_{\eta_c}^2 m_{\eta_c}^4 (m_{X_2}^2-m_{\eta_c}^2)\, G_{X_2\eta_c \eta_c}  }{6m_c^2m_{X_2}^2(m_{X_2}^2-p^{\prime2})(m_{\eta_c}^2-p^2)(m_{\eta_c}^2-q^2)}\,p_\alpha q_\beta \,p\cdot q + \cdots\, , \nonumber\\
&=&\Pi_{6}(p^{\prime2},p^2,q^2)\left(-p_{\alpha}q_{\beta}\right) \,p\cdot q + \cdots\, ,
\end{eqnarray}

\begin{eqnarray}\label{Hadron-CT-7}
\Pi^7_{\mu\nu\alpha\beta}(p,q)&=&- \frac{\lambda_{X_2} f_{J/\psi}^2 m_{J/\psi}^2 \, G_{X_2J/\psi J/\psi}  }{2(m_{X_2}^2-p^{\prime2})(m_{J/\psi}^2-p^2)(m_{J/\psi}^2-q^2)}\,
\left( g_{\mu\alpha}g_{\nu\beta}+g_{\mu\beta}g_{\nu\alpha}\right) + \cdots\, , \nonumber\\
&=&\Pi_{7}(p^{\prime2},p^2,q^2)\left( -g_{\mu\alpha}g_{\nu\beta}-g_{\mu\beta}g_{\nu\alpha}\right)  + \cdots\, ,
\end{eqnarray}
where the projector
\begin{eqnarray}
P_{A}^{\mu\nu\alpha\beta}(p)&=&\frac{1}{6}\left( g^{\mu\alpha}-\frac{p^\mu p^\alpha}{p^2}\right)\left( g^{\nu\beta}-\frac{p^\nu p^\beta}{p^2}\right)\, .
\end{eqnarray}
The decay constants or pole residues are defined by,
\begin{eqnarray}
\langle0|J^{\eta_c}(0)|\eta_c(p)\rangle&=&\frac{f_{\eta_c} m_{\eta_c}^2}{2m_c}  \,\, , \nonumber \\
\langle0|J_{\mu}^{J/\psi}(0)|J/\psi(p)\rangle&=&f_{J/\psi} m_{J/\psi} \,\xi_\mu \,\, , \nonumber \\
\langle0|J_{\mu\nu}^{h_c}(0)|h_c(p)\rangle&=&f_{h_c} \epsilon_{\mu\nu\alpha\beta}\, p^\alpha \xi^\beta \,\, , \nonumber \\
\langle0|J_{\mu}^{\chi_c}(0)|\chi_c(p)\rangle&=&f_{\chi_c} m_{\chi_c}\, \zeta_\mu \,\, ,
\end{eqnarray}
\begin{eqnarray}
 \langle 0|J^0 (0)|X_0 (p)\rangle &=& \lambda_{X_0}     \, , \nonumber\\
  \langle 0|J^1_{\mu\nu}(0)|X_1(p)\rangle &=& \tilde{\lambda}_{X_1} \, \epsilon_{\mu\nu\alpha\beta} \, \varepsilon^{\alpha}p^{\beta}\, , \nonumber\\
   \langle 0|J^2_{\mu\nu}(0)|X_2 (p)\rangle &=& \lambda_{X_2} \, \varepsilon_{\mu\nu}   \, ,
\end{eqnarray}
$\tilde{\lambda}_{X_1}m_{X_1}=\lambda_{X_1}$,
and the hadronic coupling constants are defined by,
\begin{eqnarray}
\langle \eta_c(p)\eta_c(q)|X_0(p^\prime)\rangle&=& i  G_{X_0\eta_c\eta_c}\, ,\nonumber \\
\langle J/\psi(p)J/\psi(q)|X_0(p^\prime)\rangle&=& i  \xi^* \cdot \xi^* \, G_{X_0J/\psi J/\psi}\, ,\nonumber \\
\langle \chi_c(p)\eta_c(q)|X_0(p^\prime)\rangle&=& -\zeta^* \cdot q \,G_{X_0\chi_c \eta_c}\, ,
\end{eqnarray}
\begin{eqnarray}
\langle J/\psi(p)\eta_c(q)|X_1(p^\prime)\rangle&=& i\xi^* \cdot \varepsilon \,G_{X_1J/\psi \eta_c}\, ,\nonumber \\
\langle h_c(p)\eta_c(q)|X_1(p^\prime)\rangle&=& \epsilon^{\lambda\tau\rho\sigma}p_{\lambda}\xi^*_{\tau}p^\prime_\rho\varepsilon_\sigma \,p\cdot q \,G_{X_1h_c \eta_c}\, ,
\end{eqnarray}
\begin{eqnarray}
\langle \eta_c(p)\eta_c(q)|X_2(p^\prime)\rangle&=& -i \varepsilon_{\mu\nu}  p^{\mu}q^{\nu} \, p\cdot q \,G_{X_2\eta_c\eta_c}\, ,\nonumber \\
\langle J/\psi(p)J/\psi(q)|X_2(p^\prime)\rangle&=& -i \varepsilon^{\alpha\beta} \xi^*_\alpha  \xi^*_\beta \, G_{X_2J/\psi J/\psi}\, ,
\end{eqnarray}
where the $\xi_\mu$, $\zeta_\mu$, $\varepsilon_{\mu}$ and $\varepsilon_{\mu\nu} $ denote the polarization vectors of the corresponding charmonium or tetraquark states.
As the currents $J_{\alpha\beta}^{h_c}(x)$ and $J^1_{\alpha\beta}(x)$ couple potentially to  the charmonia/tetraquarks with both the quantum numbers $J^{PC}=1^{+-}$ and $1^{--}$, we introduce the projector $P_{A}^{\mu\nu\alpha\beta}(p)$ to project out the states with the $J^{PC}=1^{+-}$ \cite{WZG-cccc-IJMPA}, and we would like to give some explanations in the Appendix.

At the hadron side,  there exists a factor,
\begin{eqnarray}
\frac{1}{m^2_{\eta_c}-q^2}\, \,\,\, {\rm or}\,\,\,\, \frac{1}{m^2_{J/\psi}-q^2}\, ,
\end{eqnarray}
in the components $\Pi_i(p^{\prime2},p^2,q^2)$ with $i=1-7$, while at the QCD side, we would obtain a pole term,
\begin{eqnarray}
\frac{1}{u-q^2}\, ,
\end{eqnarray}
with $u\geq 4m_c^2$. 
If we could take  the chiral limit $m^2_{\eta_c} \to 0$, $m^2_{J/\psi}\to 0$ and $u \to 0$, we expect to match the hadron side with the QCD side in the limit $q^2 \to 0$ with respect to a pole,
\begin{eqnarray}
\frac{1}{q^2}\, ,
\end{eqnarray}
at both sides, therefore we could only retain the ground state contribution as a good approximation. 
In the  case of the current $j_5(x)=\bar{u}(x)i\gamma_5 u(x)-\bar{d}(x)i\gamma_5 d(x)$, the ground state and first radial excitation are the $\pi$ and $\pi(1300)$, respectively, the energy gap is very large, we could take the chiral limit and neglect the excited states, see Sect.{\bf 5.3} in Ref.\cite{Reinders85}.
 However, in the present case, the masses of the $\eta_c$, $J/\psi$, $\eta_c^\prime$ and $\psi^\prime$ are of the same order, we cannot take the chiral limit, and have to resort other trick to match the two sides.

It is straightforward to obtain the hadronic spectral densities $\rho_H(s^\prime,s,u)$ through triple dispersion relation,
\begin{eqnarray}\label{dispersion-3}
\Pi_{H}(p^{\prime2},p^2,q^2)&=&\int_{16m_c^2}^\infty ds^{\prime} \int_{4m_c^2}^\infty ds \int_{4m_c^2}^\infty du \frac{\rho_{H}(s^\prime,s,u)}{(s^\prime-p^{\prime2})(s-p^2)(u-q^2)}\, ,
\end{eqnarray}
where
\begin{eqnarray}
\rho_{H}(s^\prime,s,u)&=&{\lim_{\epsilon_3\to 0}}\,\,{\lim_{\epsilon_2\to 0}} \,\,{\lim_{\epsilon_1\to 0}}\,\,\frac{ {\rm Im}_{s^\prime}\, {\rm Im}_{s}\,{\rm Im}_{u}\,\Pi_{H}(s^\prime+i\epsilon_3,s+i\epsilon_2,u+i\epsilon_1) }{\pi^3} \, ,
\end{eqnarray}
and the subscript $H$ stands for the hadron side. According the discussions in Refs.\cite{WZG-Landao,WZG-local}, the four-quark currents $J^0(x)$, $J^1_{\alpha\beta}(x)$ and $J^2_{\alpha\beta}(x)$ are local currents, and couple potentially to the tetraquark states, not the two-meson scattering states.
Although the variables $p^\prime$, $p$ and $q$ obey  conservation of the momentum $p^\prime=p+q$, we can obtain a nonzero imaginary part for all the variables $p^{\prime2}$, $p^2$ and $q^2$ by taking the $p^{\prime2}$, $p^2$ and $q^2$ as free parameters to determine the spectral densities.

At the QCD side, we contract all the quark fields with the Wick's theorem and take account of the perturbative terms and gluon condensate contributions in the operator product expansion, as the three-gluon condensate contributions are depressed by additional inverse powers of Borel parameters and play a tiny role. Then we can obtain the QCD spectral densities of the components $\Pi_{i}(p^{\prime2},p^2,q^2)$ with $i=1-7$ through double dispersion relation directly,
\begin{eqnarray}\label{dispersion-2}
\Pi_{QCD}(p^{\prime2},p^2,q^2)&=& \int_{4m_c^2}^\infty ds \int_{4m_c^2}^\infty du \frac{\rho_{QCD}(p^{\prime2},s,u)}{(s-p^2)(u-q^2)}\, ,
\end{eqnarray}
as
\begin{eqnarray}
{\rm lim}_{\epsilon \to 0}{\rm Im}\,\Pi_{QCD}(s^\prime+i\epsilon_3,p^2,q^2)&=&0\, ,
\end{eqnarray}
Naively, we expect to obtain the triple dispersion relation,
\begin{eqnarray}\label{dispersion-3-QCD}
\Pi_{QCD}(p^{\prime2},p^2,q^2)&=&\int_{16m_c^2}^\infty ds^{\prime} \int_{4m_c^2}^\infty ds \int_{4m_c^2}^\infty du \frac{\rho_{QCD}(s^\prime,s,u)}{(s^\prime-p^{\prime2})(s-p^2)(u-q^2)}\, ,
\end{eqnarray}
to match with the hadron side, $\Pi_{H}(p^{\prime2},p^2,q^2)=\Pi_{QCD}(p^{\prime2},p^2,q^2)$, see Eq.\eqref{dispersion-3}.

The triple dispersion relation in Eq.\eqref{dispersion-3} at the hadron side cannot match with the double dispersion relation in Eq.\eqref{dispersion-2} at the QCD side,
therefore  we have to match the hadron side with the QCD side of the correlation functions according to the rigorous quark-hadron duality suggested in Refs.\cite{WZG-ZJX-Zc-Decay,WZG-Y4660-Decay},
\begin{eqnarray}\label{Duality}
\int_{4m_c^2}^{s_0}ds \int_{4m_c^2}^{u_0}du  \left[ \int_{16m_c^2}^{\infty}ds^\prime  \frac{\rho_H(s^\prime,s,u)}{(s^\prime-p^{\prime2})(s-p^2)(u-q^2)} \right] &=&\int_{4m_c^2}^{s_{0}}ds \int_{4m_c^2}^{u_0}du  \frac{\rho_{QCD}(s,u)}{(s-p^2)(u-q^2)} \, ,   \nonumber\\
\end{eqnarray}
and accomplish the integral over $ds^\prime$ firstly at the hadron side. As the higher resonances and continuum states in the $s^\prime$ channels are unclear, letting alone their transitions to the ground state meson pairs,
we introduce the free parameters $C_{i}$ with $i=1-7$ to stand for the contributions concerning  the higher resonances and continuum states in the $s^\prime$ channel. For example,
\begin{eqnarray}
C_1&=&\int_{s^\prime_0}^\infty ds^\prime \frac{\tilde{\rho}_{H}(s^\prime,m_{\eta_c}^2,m_{\eta_c}^2)}{s^\prime-p^{\prime2 }}\, ,
\end{eqnarray}
where $\rho_{H}(s^\prime,m_{\eta_c}^2,m_{\eta_c}^2)=\tilde{\rho}_{H}(s^\prime,s,u)
\delta(s-m_{\eta_c}^2)\delta(u-m_{\eta_c}^2)$.
We suggest such a scheme in Refs.\cite{WZG-ZJX-Zc-Decay,WZG-Y4660-Decay} as a conjecture, direct applications 
indicate such a scheme works  well.

Afterwards, we would like to present the hadron representations clearly,
\begin{eqnarray}\label{HS-CT-1}
\Pi_{1}(p^{\prime2},p^2,q^2)&=& \frac{\lambda_{X_0} f_{\eta_c}^2 m_{\eta_c}^4 G_{X_0\eta_c \eta_c} }{4m_c^2(m_{X_0}^2-p^{\prime2})(m_{\eta_c}^2-p^2)(m_{\eta_c}^2-q^2)}+\frac{C_1 }{(m_{\eta_c}^2-p^2)(m_{\eta_c}^2-q^2)} \, ,
\end{eqnarray}

\begin{eqnarray}\label{HS-CT-2}
\Pi_{2}(p^{\prime2},p^2,q^2)&=& \frac{\lambda_{X_0} f_{J/\psi}^2 m_{J/\psi}^2 G_{X_0J/\psi J/\psi} }{(m_{X_0}^2-p^{\prime2})(m_{J/\psi}^2-p^2)(m_{J/\psi}^2-q^2)} +\frac{C_2 }{(m_{J/\psi}^2-p^2)(m_{J/\psi}^2-q^2)} \, ,
\end{eqnarray}

\begin{eqnarray}\label{HS-CT-3}
\Pi_{3}(p^{\prime2},p^2,q^2)&=& \frac{\lambda_{X_0} f_{\chi_c} m_{\chi_c} f_{\eta_c}m_{\eta_c}^2 G_{X_0\chi_c \eta_c}   }{2m_c(m_{X_0}^2-p^{\prime2})(m_{\chi_c}^2-p^2)(m_{\eta_c}^2-q^2)} + \frac{C_3}{(m_{\chi_c}^2-p^2)(m_{\eta_c}^2-q^2)}\, ,
\end{eqnarray}

\begin{eqnarray}\label{HS-CT-4}
\Pi_4(p^{\prime2},p^2,q^2)&=&\frac{\tilde{\lambda}_{X_1} f_{J/\psi}m^3_{J/\psi}f_{\eta_c} m_{\eta_c}^2 G_{X_1J/\psi \eta_c} }{2m_c(m_{X_1}^2-p^{\prime2})(m_{J/\psi}^2-p^2)(m_{\eta_c}^2-q^2)} + \frac{C_4}{(m_{J/\psi}^2-p^2)(m_{\eta_c}^2-q^2)}\, ,
\end{eqnarray}

\begin{eqnarray}\label{HS-CT-5}
\Pi_5(p^{\prime2},p^2,q^2)&=&\frac{\tilde{\lambda}_{X_1} f_{h_c}m^2_{h_c}f_{\eta_c} m_{\eta_c}^4 G_{X_1h_c \eta_c} }{9m_c(m_{X_1}^2-p^{\prime2})(m_{h_c}^2-p^2)(m_{\eta_c}^2-q^2)} +\frac{C_5}{(m_{h_c}^2-p^2)(m_{\eta_c}^2-q^2)} \, ,
\end{eqnarray}

\begin{eqnarray}\label{HS-CT-6}
\Pi_{6}(p^{\prime2},p^2,q^2)&=& \frac{\lambda_{X_2} f_{\eta_c}^2 m_{\eta_c}^4 (m_{X_2}^2-m_{\eta_c}^2)\, G_{X_2\eta_c \eta_c}  }{6m_c^2m_{X_2}^2(m_{X_2}^2-p^{\prime2})(m_{\eta_c}^2-p^2)(m_{\eta_c}^2-q^2)} + \frac{C_6}{(m_{\eta_c}^2-p^2)(m_{\eta_c}^2-q^2)} \, ,
\end{eqnarray}

\begin{eqnarray}\label{HS-CT-7}
\Pi_{7}(p^{\prime2},p^2,q^2)&=& \frac{\lambda_{X_2} f_{J/\psi}^2 m_{J/\psi}^2 \, G_{X_2J/\psi J/\psi}  }{2(m_{X_2}^2-p^{\prime2})(m_{J/\psi}^2-p^2)(m_{J/\psi}^2-q^2)}+ \frac{C_7}{(m_{J/\psi}^2-p^2)(m_{J/\psi}^2-q^2)}\, .
\end{eqnarray}

The variables $p^{\prime2}$ in Eq.\eqref{Duality} can be set as $p^{\prime2}=\alpha p^2$, where the $\alpha$ is a constant. According to the mass poles at $s^\prime=m^2_{X_{0,1,2}}$ and $s=m^2_{\eta_c,J/\psi,\chi_c,h_c}$, we can obtain  an approximated relation $s^\prime=4s$, therefore, we can estimate that it is reasonable to set  $\alpha=1\sim4$ in the present work. It is just a phenomenological  trick, not really $p^{\prime2}=(1\sim4) p^2$, as the $p^{\prime2}$, $p^2$ and $q^2$ are free parameters after performing the operator product expansion. In numerical calculations, we obtain the optimal value $\alpha=2$ in all the QCD sum rules via trial and error, which is consistent with our previous studies \cite{WZG-YXS-cccc-AAPPS}.

Then we perform double Borel transform in regard to variables $P^2=-p^2$ and $Q^2=-q^2$ and set the double Borel parameters as $T_1^2=T_2^2=T^2$  to expect appearance of flat Borel platforms, which is one  criterion of the QCD sum rules. Finally, we  get seven QCD sum rules, as an example,
\begin{eqnarray}\label{QCDSR-X0eta-eta}
&&\frac{\lambda_{X_0\eta_c \eta_c} G_{X_0\eta_c\eta_c}}{2 (\widetilde{m}^2_{X_0}-m^2_{\eta_c})}\left[\exp\left(-\frac{m^2_{\eta_c}}{T^2}\right) -\exp\left(-\frac{\widetilde{m}^2_{X_0}}{T^2}\right) \right] \exp\left(-\frac{m^2_{\eta_c}}{T^2}\right)+C_{1} \exp\left(-\frac{2m^2_{\eta_c}}{T^2}\right) \nonumber\\
&=&-\frac{3}{64\pi^4} \int^{s_{\eta_c}^0}_{4m^2_c} ds \int^{s_{\eta_c}^0}_{4m^2_c} du  \, su(s+u-4m^2_c)\sqrt{1-\frac{4m^2_c}{s}} \sqrt{1-\frac{4m^2_c}{u}} \exp\left(-\frac{s+u}{T^2}\right) \nonumber\\
&&+\frac{ m^2_c }{144\pi^2}\langle\frac{\alpha_{s}GG}{\pi}\rangle \int^{s_{\eta_c}^0}_{4m^2_c} ds \int^{s_{\eta_c}^0}_{4m^2_c} du \sqrt{1-\frac{4m^2_c}{u}} \exp\left(-\frac{s+u}{T^2}\right)\nonumber\\
&&\frac{s \left[-6s^3u+s m^2_c(-3s^2+58su+36u^2)+8m^4_c(4s^2-23su-9u^2)+m^6_c(180u-68s)\right]} {\sqrt{s\left(s-4m^2_c\right)}^5} \nonumber\\
&&+\frac{1}{576\pi^2}\langle\frac{\alpha_{s}GG}{\pi}\rangle \int^{s_{\eta_c}^0}_{4m^2_c} ds \int^{s_{\eta_c}^0}_{4m^2_c} du \exp\left(-\frac{s+u}{T^2}\right) \nonumber\\
&&\frac{-27su(s+u)+4m^2_c(7s^2+36su+7u^2)-32m^4_c(s+u)-112m^6_c} {\sqrt{s\left(s-4m^2_c\right)} \sqrt{u\left(u-4m^2_c\right)}} \nonumber\\
&&-\frac{1}{16\pi^2}\langle\frac{\alpha_{s}GG}{\pi}\rangle \int^{s_{\eta_c}^0}_{4m^2_c} ds \int^{s_{\eta_c}^0}_{4m^2_c} du \frac{ (s-m^2_c) \left[s^2-2m^2_c(3s+u)+8m^4_c\right]} { \sqrt{s\left(s-4m^2_c\right)}^3} \sqrt{u\left(u-4m^2_c\right)} \exp\left(-\frac{s+u}{T^2}\right)\, , \nonumber\\
&&
\end{eqnarray}
where the notation is defined by,
\begin{eqnarray}
\lambda_{X_0\eta_c\eta_c}&=&\frac{\lambda_{X_0} f^2_{\eta_c} m^4_{\eta_c}}{4m^2_c}\, .
\end{eqnarray}
We neglect the other six QCD sum rules for simplicity and readers can get them by contacting us via email.
In numerical calculations, we suppose  the $C_{i}$ as unknown parameters and search for the suitable values in order to obtain flat Borel platforms for the hadronic coupling constants via trial and error \cite{WZG-ZJX-Zc-Decay,WZG-Y4660-Decay,WZG-X4140-decay,WZG-Zcs3985-decay,WZG-Zcs4123-decay,
WZG-Z4500-LC-decay,WZG-3872-decay,WZG-Y4500-decay}. It is just a assumption and should be examined by the experimental data to see whether or not it is feasible.
In details, there appear endpoint divergences at the thresholds  $s=4m_c^2$ and $u=4m_c^2$ by the factors $s-4m_c^2$ and $u-4m_c^2$ in the denominators. The routine replacements $s-4m_c^2\to s-4m_c^2+\Delta^2$ and $u-4m_c^2\to u-4m_c^2+\Delta^2$ with $\Delta^2=m_c^2$ are performed to regularize the divergences due to the tiny contributions of the gluon condensates \cite{WZG-YXS-cccc-AAPPS,WZG-5c-NPB,WZG-6c-IJMPA}.

\section{Numerical results and discussions}	
At the QCD side, we take the standard gluon condensate $\langle \frac{\alpha_s GG}{\pi}\rangle=0.012\pm0.004\,\rm{GeV}^4$
\cite{SVZ79,Reinders85,Colangelo-Review} and the $\overline{MS}$ mass $m_{c}(m_c)=(1.275\pm0.025)\,\rm{GeV}$  from the Particle Data Group \cite{PDG}.
In addition, we also allow for the energy-scale dependence of the $\overline{MS}$ mass,
\begin{eqnarray}
m_c(\mu)&=&m_c(m_c)\left[\frac{\alpha_{s}(\mu)}{\alpha_{s}(m_c)}\right]^{\frac{12}{33-2n_f}} \, ,\nonumber\\
\alpha_s(\mu)&=&\frac{1}{b_0t}\left[1-\frac{b_1}{b_0^2}\frac{\log t}{t} +\frac{b_1^2(\log^2{t}-\log{t}-1)+b_0b_2}{b_0^4t^2}\right]\, ,
\end{eqnarray}
where $t=\log \frac{\mu^2}{\Lambda^2}$, $b_0=\frac{33-2n_f}{12\pi}$, $b_1=\frac{153-19n_f}{24\pi^2}$, $b_2=\frac{2857-\frac{5033}{9}n_f+\frac{325}{27}n_f^2}{128\pi^3}$,  $\Lambda=213\,\rm{MeV}$, $296\,\rm{MeV}$ and $339\,\rm{MeV}$ for the flavors $n_f=5$, $4$ and $3$, respectively  \cite{PDG}.
In the present work, the flavor number is set as $n_f=4$ to study the fully-charm tetraquark states.

At the hadron side, the parameters are taken as $m_{\eta_c}=2.9834\,\rm{GeV}$,  $m_{J/\psi}=3.0969\,\rm{GeV}$, $m_{h_c}=3.525\,\rm{GeV}$,  $m_{\chi_{c1}}=3.51067\,\rm{GeV}$ from the Particle Data Group \cite{PDG},
$s^0_{h_c}=(3.9\,\rm{GeV})^2$, $s^0_{\chi_{c1}}=(3.9\,\rm{GeV})^2$, $s^0_{J/\psi}=(3.6\,\rm{GeV})^2$, $s^0_{\eta_c}=(3.5\,\rm{GeV})^2$,
 $f_{h_c}=0.235\,\rm{GeV}$, $f_{J/\psi}=0.418 \,\rm{GeV}$, $f_{\eta_c}=0.387 \,\rm{GeV}$   \cite{Becirevic}, $f_{\chi_{c1}}=0.338\,\rm{GeV}$ \cite{Charmonium-PRT}, $m_{X_0}=6.52\,\rm{GeV}$, $\lambda_{X_0}=6.17\times 10^{-1}\,\rm{GeV}^5$, $m_{X_1}=6.57\,\rm{GeV}$, $\lambda_{X_1}=5.17\times 10^{-1}\,\rm{GeV}^5$,      $m_{X_2}=6.60\,\rm{GeV}$, $\lambda_{X_2}=7.95\times 10^{-1}\,\rm{GeV}^5$ from the QCD sum rules \cite{WZG-cccc-IJMPA}.

At beginning,  we set the free parameters as $C_i=0$, but we cannot obtain any stable platform, which indicates that the contributions concerning the higher resonances and continuum states are considerable. Therefore, we try to obtain stable platforms by varying the values of the unknown parameters  $C_i$ via trial and error. At last, we obtain the values,
\begin{eqnarray}
C_{1}&=&0.26\,T^2\,\rm{GeV}^8\, ,\nonumber\\
C_{2}&=&0.0073\,T^2\,\rm{GeV}^8\, ,\nonumber\\
C_{3}&=&0.034\,T^2\,\rm{GeV}^7\, ,\nonumber\\
C_4&=&0.007\,T^2\,{\rm GeV}^9\, ,\nonumber\\
C_5&=&0.0034\,T^2\,{\rm GeV}^6\, ,\nonumber\\
C_6&=&0.0032\,T^2\,{\rm GeV}^4\, ,\nonumber\\
C_7&=&0.007\,T^2\,{\rm GeV}^8\, ,
\end{eqnarray}
which could lead to the Borel platforms,
\begin{eqnarray}
T^2_{X_0\eta_c\eta_c}&=&(3.0-4.0)\,\rm{GeV}^2\, ,\nonumber\\
T^2_{X_0J/\psi J/\psi}&=&(1.9-2.9)\,\rm{GeV}^2\, ,\nonumber\\
T^2_{X_0\chi_c \eta_c}&=&(3.2-4.2)\,\rm{GeV}^2\, ,\nonumber\\
T^2_{X_1 J/\psi \eta_c}&=&(3.8-4.8)\,{\rm GeV}^2\, ,\nonumber\\
T^2_{X_1 h_c \eta_c}&=&(2.9-3.9)\,{\rm GeV}^2\, ,\nonumber\\
T^2_{X_2 \eta_c \eta_c}&=&(2.2-3.2)\,{\rm GeV}^2\, ,\nonumber\\
T^2_{X_2 J/\psi J/\psi}&=&(2.4-3.4)\,{\rm GeV}^2\, ,
\end{eqnarray}
where the subscripts $X_0\eta_c \eta_c$, $X_0J/\psi J/\psi$,
$X_0\chi_c \eta_c$, $X_1 J/\psi\eta_c$, $X_1 h_c\eta_c$, $X_2\eta_c \eta_c$ and
$X_2J/\psi J/\psi$ denote the corresponding channels (modes).
Therefore, we obtain seven flat platforms with uniform intervals  $T^2_{max}-T^2_{min}=1\,\rm{GeV}^2$, just like in our previous works \cite{WZG-ZJX-Zc-Decay,WZG-Y4660-Decay,WZG-X4140-decay,WZG-Zcs3985-decay,WZG-Zcs4123-decay,
WZG-Z4500-LC-decay,WZG-3872-decay,WZG-Y4500-decay}, where the max and min denote the maximum and minimum, respectively.

Before analyzing the numerical results, it is crucial to establish the uncertainties of the hadronic coupling constants. The uncertainties  not only originate from the decay constants (or pole residues), but also originate from the parameters at the QCD side,  we should  avoid overestimating the uncertainties. In details, the uncertainties of the channel $X_0 \rightarrow J/\psi + J/\psi$, for example, are presented as $\lambda_{X_0}f_{J/\psi}^2G_{X_0J/\psi J/\psi} = \bar{\lambda}_{X_0}\bar{f}_{J/\psi}^2\bar{G}_{X_0J/\psi J/\psi}
+\delta\,\lambda_{X_0}f_{J/\psi}^2G_{X_0J/\psi J/\psi}$, $C_{2} = \bar{C}_{2}+\delta C_{2}$,
\begin{eqnarray}\label{Uncertainty-4}
\delta\,\lambda_{X_0}f_{J/\psi}^2G_{X_0J/\psi J/\psi} &=&\bar{\lambda}_{X_0}\bar{f}_{J/\psi}^2\bar{G}_{X_0J/\psi J/\psi}\left( 2\frac{\delta f_{J/\psi}}{\bar{f}_{J/\psi}} +\frac{\delta \lambda_{X_0}}{\bar{\lambda}_{X_0}}+\frac{\delta G_{X_0J/\psi J/\psi}}{\bar{G}_{X_0J/\psi J/\psi}}\right)\, ,
\end{eqnarray}
where the short overline $\bar{}$ denotes the central value. Then we approximately set $\delta C_{2}=0$ and $\frac{\delta f_{J/\psi}}{\bar{f}_{J/\psi}} = \frac{\delta \lambda_{X_0}}{\bar{\lambda}_{X_0}}=\frac{\delta G_{X_0J/\psi J/\psi}}{\bar{G}_{XJ/\psi J/\psi}}$ to get the uncertainties of the hadronic coupling constants.

Finally, we obtain the values of the hadronic coupling constants,
\begin{eqnarray} \label{HCC-values}
G_{X_0\eta_c \eta_c} &=&7.49^{+3.59}_{-3.67}\,\rm{GeV}^{3} \, , \nonumber\\
G_{X_0J/\psi J/\psi} &=&0.35^{+0.09}_{-0.07}\,\rm{GeV}^{3} \, , \nonumber\\
G_{X_0\chi_c \eta_c} &=&3.31^{+0.66}_{-0.59}\,{\rm GeV}^{2} \, , \nonumber\\
G_{X_1 J/\psi \eta_c}&=& 0.14^{+0.26}_{-0.14}\,{\rm GeV}^{3} \, ,\nonumber\\
G_{X_1 h_c \eta_c}&=& 0.32^{+0.08}_{-0.08}\,{\rm GeV}^{-3} \, ,\nonumber\\
G_{X_2 \eta_c \eta_c}&=& 0.20^{+0.07}_{-0.07}\,{\rm GeV}^{-3} \, ,\nonumber\\
G_{X_2 J/\psi J/\psi}&=& 0.59^{+0.15}_{-0.13}\,{\rm GeV}^{3} \, .
\end{eqnarray}

\begin{figure}
 \centering
  \includegraphics[totalheight=5cm,width=7cm]{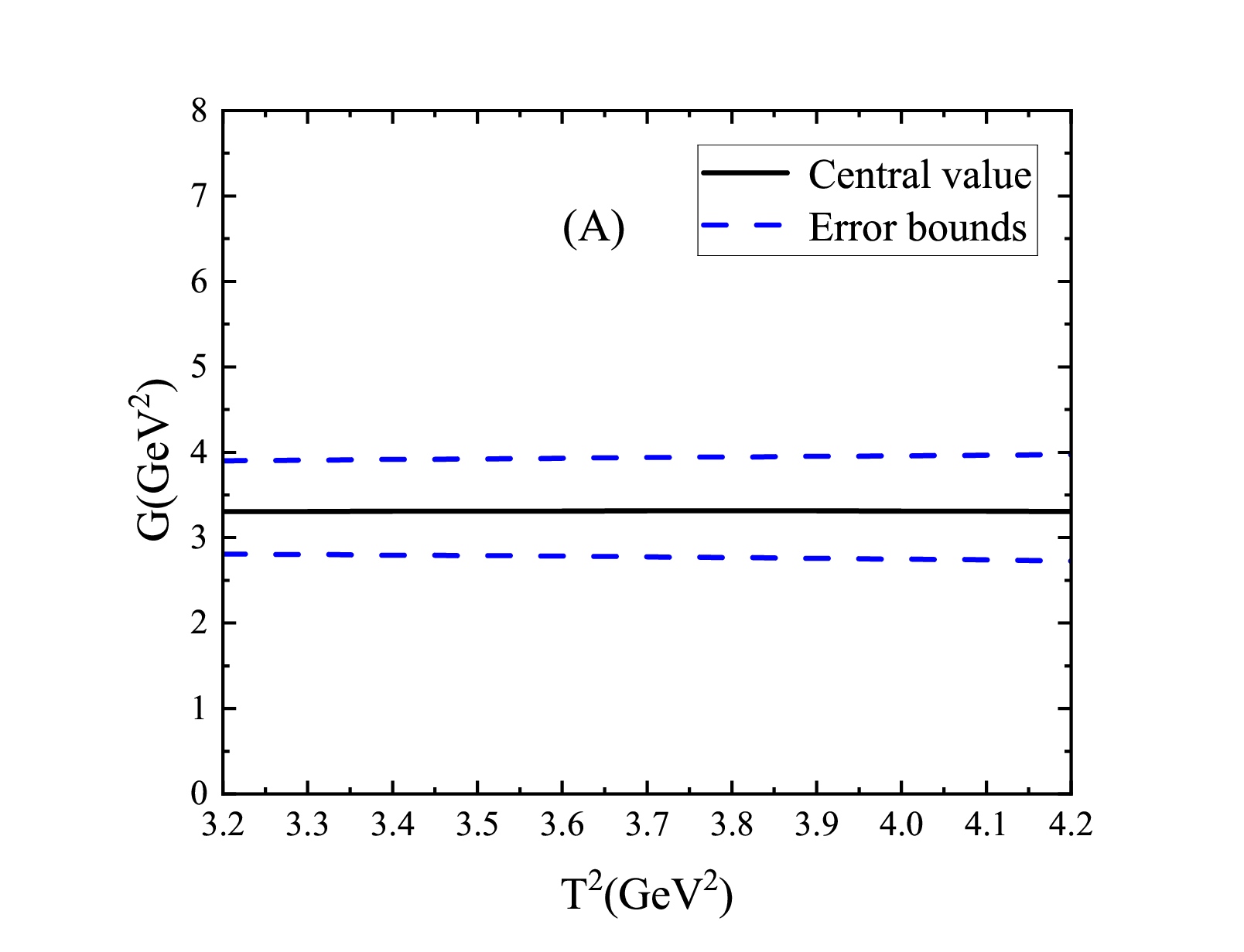}
  \includegraphics[totalheight=5cm,width=7cm]{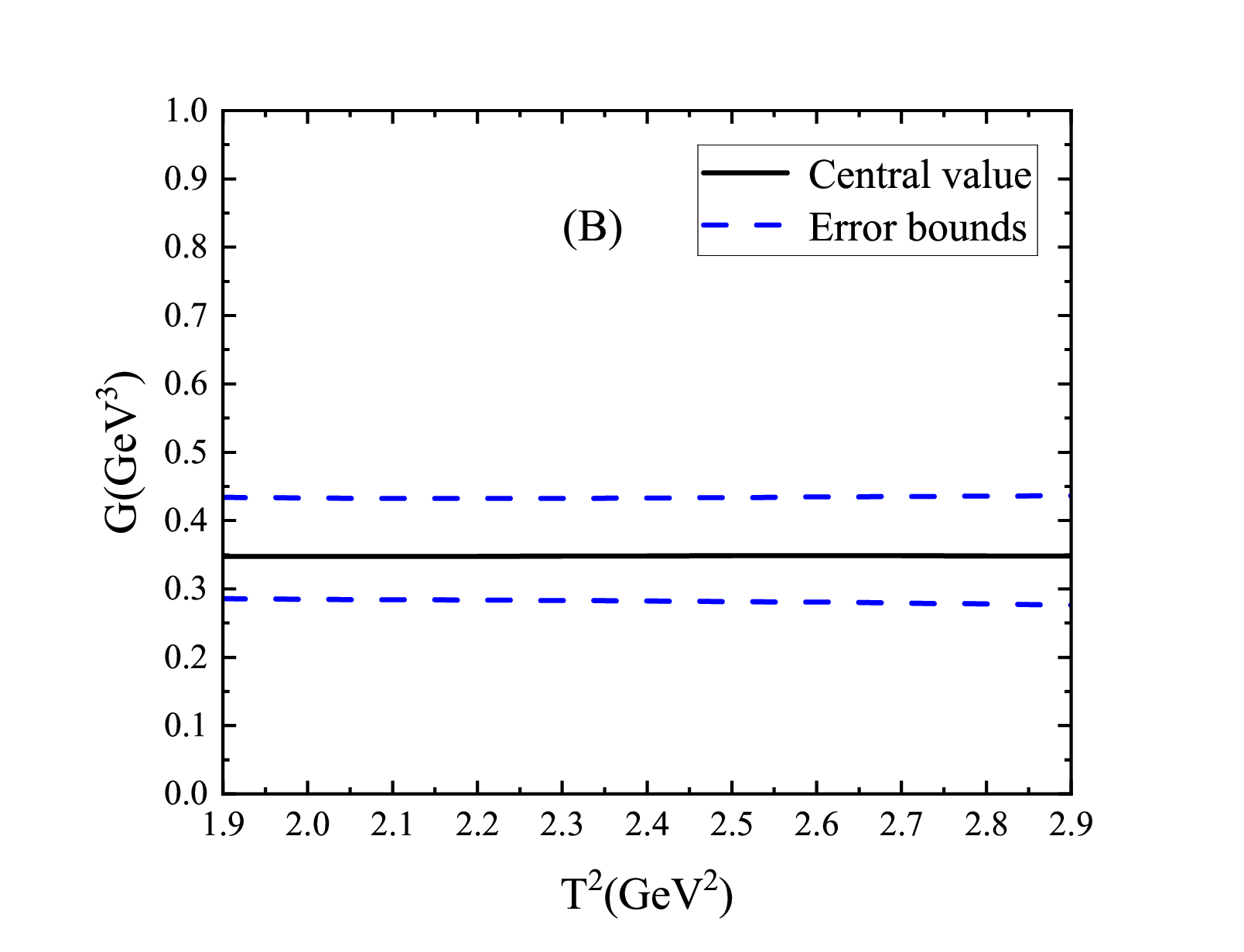}
        \caption{The central values of the hadronic coupling constants with variations of the  Borel parameters $T^2$, where the (A) and (B)  denote the $G_{X_0 \eta_c \chi_c}$  and $G_{X_0 J/\psi J/\psi}$,  respectively.}\label{fig-chi-eta-diJpsi}
\end{figure}

In Fig.\ref{fig-chi-eta-diJpsi}, the curves of the hadronic coupling constants $G_{X_0\chi_c\eta_c}$ and $G_{X_0 J/\psi J/\psi}$ with variations of the Borel parameters $T^2$ are plotted in the Borel windows as an example. There appear flat platforms clearly, and thus we can extract the hadronic coupling constants reasonably.

Then we take the masses $m_{X_0}=6.52\,\rm{GeV}$,  $m_{X_1}=6.57\,\rm{GeV}$, and $m_{X_2}=6.60\,\rm{GeV}$ obtained from the QCD sum rules \cite{WZG-cccc-IJMPA}, and get the partial decay widths directly,
\begin{eqnarray} \label{X0-decay}
\Gamma(X_0\to \eta_c \eta_c)&=& 68.94^{+81.93}_{-51.01} \,\rm{MeV}\, , \nonumber\\
\Gamma(X_0\to J/\psi J/\psi)&=& 0.41^{+0.23}_{-0.15} \,\rm{MeV}\, , \nonumber\\
\Gamma(X_0\to \chi_{c1} \eta_c)&=& 0.83^{+0.37}_{-0.27} \,\rm{MeV}\, ,
\end{eqnarray}

\begin{eqnarray}\label{X1-decay}
\Gamma(X_1 \to J/\psi \eta_c)&=& 0.024^{+0.17}_{-0.024}\, {\rm MeV}\, ,\nonumber\\
\Gamma(X_1 \to h_c \eta_c)&=& 28.35^{+15.95}_{-12.40}\,{\rm MeV}\, ,
\end{eqnarray}

\begin{eqnarray}\label{X2-decay}
\Gamma(X_2 \to \eta_c \eta_c)&=& 4.49^{+3.69}_{-2.60}\, {\rm MeV}\, ,\nonumber\\
\Gamma(X_2 \to J/\psi J/\psi)&=& 0.40^{+0.22}_{-0.16}\, {\rm MeV}\, ,
\end{eqnarray}
and therefore the full widths,
\begin{eqnarray}
\Gamma_{X_0}&=&70.18^{+82.53}_{-51.43}\,\rm{MeV}\, ,
\end{eqnarray}
\begin{eqnarray}
\Gamma_{X_1}&=& 28.37^{+16.12}_{-12.42}\, {\rm MeV}\, ,
\end{eqnarray}
\begin{eqnarray}
\Gamma_{X_2}&=& 4.89^{+3.91}_{-2.76}\, {\rm MeV}\, .
\end{eqnarray}

In our previous studies, the mass spectrum of the ground states and first/second/third radial excitations and the decay widths of the ground states and first radial excitations of the $A\bar{A}$-type fully-charm tetraquark states are studied via the QCD sum rules, and the results show that the $X(6552)$ can be assigned as the first radial excitation of the  $A\bar{A}$-type scalar tetraquark state considering  both the masses and decay widths \cite{WZG-cccc-NPB,WZG-YXS-cccc-AAPPS}.
In Ref.\cite{WZG-cccc-IJMPA}, we studied the mass spectrum of the ground state $\widetilde{V}\bar{\widetilde{V}}$-type scalar, axialvector and tensor fully-charm tetraquark states with the QCD sum rules. The numerical results indicate that the ground state $\widetilde{V}\overline{\widetilde{V}}$-type tetraquark states and the first radial excited states of the $A\bar{A}$-type tetraquark states have almost degenerated masses, about $0.35\pm0.09\,\rm{GeV}$ above the $J/\psi J/\psi$ threshold.

In the present work, the predicted width $\Gamma_{X_0}=70.18^{+82.53}_{-51.43}\,\rm{MeV}$ is compatible with the experimental data $\Gamma_{X(6552)} = 124^{+32}_{-26}\pm33 \,\rm{ MeV}$ from the CMS collaboration \cite{CMS-cccc-2023} within the range of uncertainties, which supports assigning the $X(6552)$ as the ground state $\widetilde{V}\overline{\widetilde{V}}$-type scalar tetraquark state, while the widths of the tetraquark states with higher spins $\Gamma_{X_1}=28.37^{+16.12}_{-12.42}\,\rm{MeV}$ and $\Gamma_{X_2}=4.89^{+3.91}_{-2.76}\,\rm{MeV}$ are too small to match with the experimental data.

As a hadron has several Fock states, the $X(6552)$ maybe have both the 1S $\widetilde{V}\overline{\widetilde{V}}$-type and 2S $A\bar{A}$-type scalar tetraquark components.
Furthermore, the relative branching ratios  are quite different from each other,
\begin{eqnarray}
\Gamma\left(X^{\widetilde{V}\overline{\widetilde{V}}}_0 \to \eta_c \eta_c:J/\psi J/\psi:\chi_{c1} \eta_c\right) &=& 1.00:0.0059:0.012 \, , \nonumber\\
\Gamma\left(X^{A\bar{A}}_0 \to \eta_c \eta_c:J/\psi J/\psi:\chi_{c1} \eta_c\right) &=& 0.066:1.00:0.0024 \, ,
\end{eqnarray}
which indicate that the main decay channels are $X_0\to \eta_c \eta_c$ for the 1S $\widetilde{V}\overline{\widetilde{V}}$-type scalar tetraquark state and $X_0\to J/\psi J/\psi$ for the 2S $A\bar{A}$-type scalar tetraquark state, respectively. There still need more experimental data to diagnose the nature of the fully-charm tetraquark states.
Other predictions are served as meaningful guides for the high energy experiments, awaiting to be examined in the future.

\section{Conclusion}
In the present study, we introduce a relative P-wave to construct the doubly-charm vector diquark, and therefore construct the scalar and tensor four-quark currents to investigate the decay widths of the fully-charm tetraquark states with the $J^{PC}=0^{++}$, $1^{+-}$ and $2^{++}$ via the QCD sum rules.
We take account of the perturbative terms and gluon condensate contributions in the operator product expansion and then match the hadron side with the QCD side based on rigorous quark-hadron duality.
The predicted width of the ground state scalar tetraquark state $\Gamma_{X_0}=70.18^{+82.53}_{-51.43}\,\rm{MeV}$ is compatible with the experimental data $\Gamma_{X(6552)} = 124^{+32}_{-26}\pm33 \,\rm{ MeV}$ from the CMS collaboration within the range of uncertainties, which supports assigning the $X(6552)$ as the ground state $\widetilde{V}\overline{\widetilde{V}}$-type scalar tetraquark state.
The relative branching ratios of the ground state $\widetilde{V}\overline{\widetilde{V}}$-type scalar tetraquark state and the first radial excitation of the $A\bar{A}$-type scalar tetraquark state are quite different, which can be used to clarify the nature of the $X(6552)$.
We also expect the other predictions will be confirmed in the future experiments.

\section*{Appendix}
For simplicity, we introduce the notation $J_{\alpha\beta}(x)$ to denote the $J_{\alpha\beta}^{h_c}(x)$ and $J^1_{\alpha\beta}(x)$, and resort to
the two-point correlation function  $\Pi_{\mu\nu\alpha\beta}(p)$,
\begin{eqnarray}
\Pi_{\mu\nu\alpha\beta}(p)&=&i\int d^4x e^{ip \cdot x} \langle0|T\left\{J_{\mu\nu}(x)J_{\alpha\beta}^{\dagger}(0)\right\}|0\rangle \, ,
\end{eqnarray}
to illustrate how to project out the pertinent tensor structures.
At the hadron side, we isolate the ground state contributions,
\begin{eqnarray}
\Pi_{\mu\nu\alpha\beta}(p)&=&\frac{\tilde{\lambda}_{ A}^2}{M_{A}^2-p^2}\left(p^2g_{\mu\alpha}g_{\nu\beta} -p^2g_{\mu\beta}g_{\nu\alpha} -g_{\mu\alpha}p_{\nu}p_{\beta}-g_{\nu\beta}p_{\mu}p_{\alpha}+g_{\mu\beta}p_{\nu}p_{\alpha}+g_{\nu\alpha}p_{\mu}p_{\beta}\right) \nonumber\\
&&+\frac{\tilde{\lambda}_{ V}^2}{M_{V}^2-p^2}\left( -g_{\mu\alpha}p_{\nu}p_{\beta}-g_{\nu\beta}p_{\mu}p_{\alpha}+g_{\mu\beta}p_{\nu}p_{\alpha}+g_{\nu\alpha}p_{\mu}p_{\beta}\right) +\cdots \, \, ,\nonumber\\
&=&\Pi_{A}(p^2)\left(p^2g_{\mu\alpha}g_{\nu\beta} -p^2g_{\mu\beta}g_{\nu\alpha} -g_{\mu\alpha}p_{\nu}p_{\beta}-g_{\nu\beta}p_{\mu}p_{\alpha}+g_{\mu\beta}p_{\nu}p_{\alpha}+g_{\nu\alpha}p_{\mu}p_{\beta}\right) \nonumber\\
&&+\Pi_{V}(p^2)\left( -g_{\mu\alpha}p_{\nu}p_{\beta}-g_{\nu\beta}p_{\mu}p_{\alpha}+g_{\mu\beta}p_{\nu}p_{\alpha}+g_{\nu\alpha}p_{\mu}p_{\beta}\right) \, ,
\end{eqnarray}
where
\begin{eqnarray}
   \langle 0|J_{\mu\nu}(0)|A(p)\rangle &=& \tilde{\lambda}_{A} \, \varepsilon_{\mu\nu\alpha\beta} \, \varepsilon^{\alpha}p^{\beta}\, , \nonumber\\
 \langle 0|J_{\mu\nu}(0)|V(p)\rangle &=& \tilde{\lambda}_{V} \left(\varepsilon_{\mu}p_{\nu}-\varepsilon_{\nu}p_{\mu} \right)\, ,
\end{eqnarray}
the   $A$ and $V$ stand for the $J^{PC}=1^{+-}$ and $1^{--}$ mesons, respectively.
We introduce the  operators $P_{A}^{\mu\nu\alpha\beta}(p)$ and $P_{V}^{\mu\nu\alpha\beta}(p)$,
\begin{eqnarray}
P_{A}^{\mu\nu\alpha\beta}(p)&=&\frac{1}{6}\left( g^{\mu\alpha}-\frac{p^\mu p^\alpha}{p^2}\right)\left( g^{\nu\beta}-\frac{p^\nu p^\beta}{p^2}\right)\, , \nonumber\\
P_{V}^{\mu\nu\alpha\beta}(p)&=&\frac{1}{6}\left( g^{\mu\alpha}-\frac{p^\mu p^\alpha}{p^2}\right)\left( g^{\nu\beta}-\frac{p^\nu p^\beta}{p^2}\right)-\frac{1}{6}g^{\mu\alpha}g^{\nu\beta}\, .
\end{eqnarray}
and project out the components $\Pi_{A}(p^2)$ and $\Pi_{V}(p^2)$ unambiguously,
\begin{eqnarray}
\widetilde{\Pi}_{A}(p^2)&=&p^2\Pi_{A}(p^2)=P_{A}^{\mu\nu\alpha\beta}(p)\Pi_{\mu\nu\alpha\beta}(p) \, , \nonumber\\
\widetilde{\Pi}_{V}(p^2)&=&p^2\Pi_{V}(p^2)=P_{V}^{\mu\nu\alpha\beta}(p)\Pi_{\mu\nu\alpha\beta}(p) \, ,
\end{eqnarray}
and
\begin{eqnarray}
P_{A}^{\mu\nu\lambda\tau}(p)\,\Pi_{\lambda\tau\alpha\beta}(p)&\propto&\Pi_{A}(p^2) \, , \nonumber\\
P_{V}^{\mu\nu\lambda\tau}(p)\,\Pi_{\lambda\tau\alpha\beta}(p)&\propto&\Pi_{V}(p^2) \, .
\end{eqnarray}
So in Eqs.\eqref{Hadron-CT-4-project}-\eqref{Hadron-CT-5-project}, we project out the contributions of the $h_c$ and  $X_1$  with the $J^{PC}=1^{+-}$ with the projectors $P_{A}^{\mu\nu\mu^\prime\nu^\prime}(p)$ and  $P_{A}^{\alpha\beta\alpha^\prime\beta^\prime}(p^\prime)$, respectively,
\begin{eqnarray}
P_{A}^{\alpha\beta\alpha^\prime\beta^\prime}(p^\prime)\,\Pi^4_{\mu\alpha\beta}(p,q)&\propto&
\widetilde{\Pi}_4(p^{\prime2},p^2,q^2)\, , \nonumber \\
P_{A}^{\mu\nu\mu^\prime\nu^\prime}(p)\,P_{A}^{\alpha\beta\alpha^\prime\beta^\prime}(p^\prime)\,
\Pi^5_{\mu\nu\alpha\beta}(p,q)&\propto&\widetilde{\Pi}_5(p^{\prime2},p^2,q^2)\, .
\end{eqnarray}

\section*{Acknowledgements}
This work is supported by National Natural Science Foundation, Grant Number 12175068, and the Fundamental Research Funds for the Central Universities.


\begin{thebibliography}{99}

\bibitem{LHCb-cccc-2006} R. Aaij et al, Sci. Bull. {\bf 65} (2020) 1983.

\bibitem{ATLAS-cccc-2023} G. Aad et al, Phys. Rev. Lett. {\bf 131} (2023)  151902.

\bibitem{CMS-cccc-2023} A. Hayrapetyan et al, Phys. Rev. Lett. {\bf132} (2024) 111901.

\bibitem{Rosner-2017} M. Karliner, J. L. Rosner and S. Nussinov,  Phys. Rev. {\bf D95} (2017)  034011.


\bibitem{Wu-2018} J. Wu, Y. R. Liu, K. Chen, X. Liu and S. L. Zhu, Phys. Rev. {\bf D97} (2018) 094015.


\bibitem{FKGuo-2018-Anwar} M. N. Anwar, J. Ferretti, F. K. Guo, E. Santopinto and B. S. Zou, Eur. Phys. J. {\bf C78} (2018)  647.

\bibitem{Polosa-2018} A. Esposito and A. D. Polosa, Eur. Phys. J. {\bf C78} (2018)  782.

\bibitem{Navarra-2019}    V. R. Debastiani and F. S. Navarra, Chin. Phys. {\bf C43} (2019)  013105.

\bibitem{Bai-2016}  Y. Bai, S. Lu and J. Osborne, 	Phys. Lett. {\bf B798} (2019) 134930.

\bibitem{Zhong-2019} M. S. Liu, Q. F. Lu, X. H. Zhong and Q. Zhao, Phys. Rev. {\bf D100} (2019)  016006.

\bibitem{Roberts-2020} M. A. Bedolla, J. Ferretti, C. D. Roberts, and E. Santopinto,   Eur. Phys. J. {\bf C80} (2020) 1004.

\bibitem{PingJL-2020} X. Jin, Y. Xue, H. Huang and J. Ping,	Eur. Phys. J. {\bf C80} (2020) 1083.

\bibitem{Zhong-2021} F. X. Liu, M. S. Liu, X. H. Zhong and Q. Zhao, Phys. Rev. {\bf D104} (2021) 116029.

\bibitem{Mutuk-2021} H. Mutuk, Eur. Phys. J. {\bf C81} (2021) 367.

\bibitem{DWC-2023-PRD} W. C. Dong and Z. G. Wang, Phys. Rev. {\bf D107} (2023), 074010.

\bibitem{YGL-2023-EPJC}  G. L. Yu, Z. Y. Li, Z. G. Wang, J. Lu and M. Yan, Eur. Phys. J. {\bf C83} (2023) 416.

\bibitem{Jia-2023} T. Q. Yan, W. X. Zhang and D. Jia, Eur. Phys. J. {\bf{C83}}, (2023) 810.

\bibitem{Ping-2024} Y. Wu, X. Liu, Y. Tan, H. Huang and J. Ping, arXiv:2403.10375 [hep-ph].

\bibitem{WZG-cccc-EPJC} Z. G. Wang, Eur. Phys. J. {\bf C77} (2017)  432.

\bibitem{Chen-2017}  W. Chen, H. X. Chen, X. Liu, T. G. Steele and S. L. Zhu, Phys. Lett. {\bf B773} (2017) 247.

\bibitem{WZG-cccc-APPB} Z. G. Wang and Z. Y. Di, Acta Phys. Polon. {\bf B50} (2019) 1335.

\bibitem{WZG-cccc-CPC} Z. G. Wang, Chin. Phys. {\bf C44} (2020)  113106.

\bibitem{WZG-cccc-IJMPA} Z. G. Wang, Int. J. Mod. Phys. {\bf A36} (2021)  2150014.


\bibitem{ZhangJR-PRD} J. R. Zhang, Phys. Rev. {\bf D103} (2021)  014018.

\bibitem{QiaoCF-2021} B. C. Yang, L. Tang, and C. F. Qiao,	Eur. Phys. J. {\bf C81} (2021) 324.
\bibitem{WZG-cccc-NPB} Z. G. Wang, Nucl. Phys. {\bf B985} (2022) 115983.


\bibitem{X6600-Azizi} S. S. Agaev, K. Azizi, B. Barsbay and H. Sundu, Phys. Lett. {\bf B844} (2023) 138089.

\bibitem{X6900-Azizi} S. S. Agaev, K. Azizi, B. Barsbay and H. Sundu, Nucl. Phys. {\bf A1041} (2024) 122768.

\bibitem{WZG-YXS-cccc-AAPPS} Z. G. Wang and X. S. Yang, AAPPS Bull. {\bf 34}
 (2024) 5.

\bibitem{Tang-2024} C. M. Tang, C. G. Duan and L. Tang, Eur. Phys. J. {\bf{C84}} (2024) 743.

\bibitem{Chen-2024} Z. Z. Chen, X. L. Chen, P. F. Yang and W. Chen, Phys. Rev. {\bf D109} (2024) 094011.

 \bibitem{Hughes-2018}  C. Hughes, E. Eichten and C. T. H. Davies, Phys. Rev. {\bf D97} (2018)  054505.

\bibitem{WangJZ-Produ-mass} J. Z. Wang, D. Y. Chen, X. Liu, and T. Matsuki,	Phys. Rev. {\bf D103} (2021) 071503.

\bibitem{Zhu-2021-NPB} R. Zhu,	Nucl. Phys. {\bf B966} (2021) 115393.

\bibitem{GuoFK-2021-PRL} X. K. Dong, V. Baru, F. K. Guo, C. Hanhart and A. Nefediev, Phys. Rev. Lett. {\bf 127} (2021) 119901.

\bibitem{XKDong-SB-2021} X. K. Dong, V. Baru, F. K. Guo, C. Hanhart and A. Nefediev, Sci. Bull. {\bf 66} (2021)  2462.


\bibitem{Gong-Zhong-2022} C. Gong, M. C. Du, Q. Zhao, X. H. Zhong and B. Zhou, Phys. Lett. {\bf B824} (2022) 136794.

\bibitem{WZG-ZJX-Zc-Decay} Z. G. Wang and J. X. Zhang,  Eur. Phys. J. {\bf C78} (2018)  14.

\bibitem{WZG-Y4660-Decay} Z. G. Wang, Eur. Phys. J. {\bf C79} (2019)  184.

\bibitem{WZG-X4140-decay}  Z. G. Wang and  Z. Y. Di,    Eur. Phys. J. {\bf C79} (2019)  72.

\bibitem{WZG-Zcs3985-decay} Z. G. Wang,  Chin. Phys. {\bf C46} (2022) 103106.


\bibitem{WZG-Zcs4123-decay} Z. G. Wang, Chin. Phys. {\bf C46} (2022)  123106.

\bibitem{WZG-Z4500-LC-decay} Z. G. Wang, Nucl. Phys. {\bf B993} (2023) 116265

\bibitem{WZG-3872-decay} Z. G. Wang, Phys. Rev. {\bf D109} (2024) 014017.

\bibitem{WZG-Y4500-decay} Z. G. Wang, Nucl. Phys. {\bf B1005} (2024) 116580.

\bibitem{SVZ79}  M. A. Shifman, A. I. Vainshtein and V. I. Zakharov, Nucl. Phys. {\bf B147} (1979) 385;
Nucl. Phys. {\bf B147} (1979) 448.

\bibitem{Reinders85} L. J. Reinders, H. Rubinstein and S. Yazaki, Phys. Rept. {\bf 127} (1985) 1.

\bibitem{WZG-Landao} Z. G. Wang, Phys. Rev. {\bf D101} (2020) 074011.

\bibitem{WZG-local} Z. G. Wang, arXiv:2102.07520.

\bibitem{WZG-5c-NPB} Z. G. Wang, Nucl. Phys. {\bf B973} (2021) 115579.

\bibitem{WZG-6c-IJMPA} Z. G. Wang, Int. J. Mod. Phys. {\bf A37} (2022) 2250166.

\bibitem{Colangelo-Review}  P. Colangelo and A. Khodjamirian, hep-ph/0010175.

\bibitem{PDG}  R. L. Workman et al, Prog. Theor. Exp. Phys. {\bf 2022} (2022) 083C01.

\bibitem{Becirevic} D. Becirevic, G. Duplancic, B. Klajn, B. Melic and F. Sanfilippo,  Nucl. Phys. {\bf B883} (2014) 306.

\bibitem{Charmonium-PRT} V. A. Novikov, L. B. Okun, M. A. Shifman, A. I. Vainshtein, M. B. Voloshin and V. I. Zakharov, Phys. Rept. {\bf 41} (1978) 1.

\end{thebibliography}
\end{document}